\documentclass[12pt,preprint]{aastex}

\newcommand{\mincir}{\raise
-2.truept\hbox{\rlap{\hbox{$\sim$}}\raise5.truept\hbox{$<$}\ }}
\newcommand{\magcir}{\raise
-2.truept\hbox{\rlap{\hbox{$\sim$}}\raise5.truept\hbox{$>$}\ }}
\newcommand{\minmag}{\raise
-2.truept\hbox{\rlap{\hbox{$<$}}\raise6.truept\hbox{$<$}\ }}

\shorttitle{X-ray number counts of normal galaxies}
\shortauthors{Georgakakis et al.}

\begin{document}

\title{X-ray number counts of normal galaxies}

\author{
A. Georgakakis\altaffilmark{1,2}, 
I. Georgantopoulos\altaffilmark{3}, 
A. Akylas\altaffilmark{3}, 
A. Zezas\altaffilmark{4},
P. Tzanavaris\altaffilmark{3}
}


\altaffiltext{1}{Astrophysics Group, Blackett Laboratory, Imperial
  College, Prince Consort Rd , London SW7 2BZ, UK}
  \email{age@imperial.ac.uk} 

\altaffiltext{2}{Marie-Curie follow}

\altaffiltext{3}{Institute of Astronomy \& Astrophysics,
  National Observatory of Athens, I. Metaxa \& B. Pavlou, Penteli,
  15236, Athens, Greece}
  
\altaffiltext{4}{Harvard-Smithsonian Center for Astrophysics, 60
  Garden Street, Cambridge, MA 02138, USA}




\begin{abstract}
We use the number counts of X-ray selected normal galaxies to explore
their evolution by combining the most recent wide-angle shallow and
pencil-beam deep samples available. The differential X-ray number
counts, $\rm dN/dS$, for early and late-type normal galaxies are
constructed separately and then compared with the predictions of the
local X-ray luminosity function under different evolution scenarios. The 
$\rm dN/dS$ of early type galaxies is consistent with no evolution out to 
$z \approx 0.5$. For late-type galaxies our analysis suggests that it
is the sources with X-ray--to--optical flux ratio $\log f_X / f_{opt}
> -2$ that are evolving the fastest. Including these systems in the 
late-type galaxy sample yields evolution of the form $\approx
(1+z)^{2.7}$ out to $z\approx 0.4$. On the contrary late-type sources
with $\log f_X /  f_{opt} < -2$ are consistent with no evolution. This
suggests that the $\log f_X / f_{opt} > -2$ population
comprises the most powerful and fast evolving starbursts at moderate
and high-$z$. We argue that although residual low-luminosity AGN
contamination may bias our results toward stronger evolution, this
is unlikely to modify our main conclusions.
\end{abstract}


\section{Introduction}
In the last few years surveys carried out by the {\it XMM-Newton} and 
{\it Chandra} missions provided the first X-ray selected samples
of normal galaxies out to cosmologically interesting redshifts
(Hornschemeier et al. 2003; Bauer et al. 2004; Georgakakis et
al. 2004; Norman et al. 2004; Georgantopoulos et al. 2005). This
development has opened the opportunity to explore the evolution at
X-ray wavelengths of the dominant population of the Universe: normal
galaxies powered by stellar processes rather than accretion on a
supermassive black hole. The unique feature of X-rays is that they are
the only tool available to directly probe the X-ray binaries and the
hot gas in galaxies, providing complementary information on their
evolution compared to other wavelengths (e.g. Ghosh \& White 2001). 

The evidence above has motivated a number of studies which use
different approaches to constrain the evolution of X-ray selected
normal galaxies. Stacking of optically selected spirals in the
{\it Chandra} Deep Field North provided the first evidence for
evolution of these systems at X-ray wavelengths (Brandt et al. 2001; 
Hornchemeier et al. 2002). A more direct method, which however
requires the detection of individual systems, is to estimate the
X-ray luminosity function at different redshift bins (e.g. Norman et
al. 2004). A simpler but equally instructive technique uses the
observed number counts in comparison with model predictions assuming
different forms of evolution (e.g. Ranalli, Comastri \&  Setti
2005). The general consensus from the complementary methods above
is that the X-ray evolution of normal galaxies to $z\approx 1-1.5$ is
similar to that obtained from other wavelengths (e.g Hopkins 2004). 

In addition to the studies above, which mainly concentrate on deep
surveys (e.g. {\it Chandra} Deep Fields) aiming to select high-$z$
normal galaxies, there have also been efforts to identify such systems
at low redshifts (Georgakakis et al. 2004; Georgantopoulos et al. 2005;
Georgakakis et al. 2006). These samples are complementary to deep
surveys, providing a tight anchor point at low-$z$ that is essential
for any conclusions on evolution. Despite the significant progress in
the field however, there is still limited number of studies that
exploit all the complementary low and high-$z$ samples available to
constrain the evolution of normal galaxies over as a wide redshift
range as possible.   


In this paper we expand on previous results
on the X-ray evolution of normal galaxies by combining the most up to
date wide-angle shallow surveys (probing on average lower-$z$) with
the latest compilation of   
normal galaxies in the {\it Chandra} Deep Field North and South (Bauer
et al. 2004). These samples are used to construct the differential
X-ray number counts of normal galaxies over about 4\,dex in flux and
to compare against different evolution models. We further improve on
previous studies by splitting the sample into early and late-type
galaxies to explore their evolution separately. This is particularly
important since different galaxy types are expected to follow
different evolution patterns. Throughout this paper we adopt $\rm
H_{o} = 70 \, km \, s^{-1} \, Mpc^{-1}$, $\rm \Omega_{M} =   0.3$ and
$\rm \Omega_{\Lambda} = 0.7$.

\section{Sample selection}\label{sample}
In this paper we combine normal galaxy samples compiled from
independent wide-area shallow and pencil-beam deep surveys spanning a 
wide range of fluxes, $f_X(\rm 0.5 - 2 \,keV) \approx 10^{-13} -
10^{-16} \, erg \, s^{-1} \, cm^{-2}$. 

At the bright flux end we use galaxies identified in (i) the Needles
in the Haystack Survey (NHS; Georgakakis et al. 2004; Georgantopoulos et
al. 2005), (ii) the 1st release of the {\it XMM-Newton} Serendipitous
Source Catalogue (1XMM) prodcued by the {\it XMM-Newton} Survey
Science Centre (see Georgakakis et al. 2006 for the galaxy selection)
and (iii) public {\it Chandra} observations, analysed as part of the
XAssist project (Ptak \& Griffiths 2003), that overlap with the Sloan
Digital Sky Survey (SDSS; Schneider et. al. 2005). For the latter
sample, full details on the field selection and normal galaxy
identification will be presented in a future publication. The galaxy
catalog for this sample is electronically available from the web
(http://www.astro.noa.gr/xray/XAssist-SDSS.txt). In brief, sources are
detected in the 0.3-8\,keV band of ACIS pointings   with exposure
times $>20$\,ks, to avoid observations that are too shallow to be
useful for our purposes. Fluxes in the 0.5-2\,keV band are estimated
from the 0.3-8\,keV count-rates in the XAssist catalogue assuming a
power-law spectral energy distribution with $\Gamma=1.8$. The
sensitivity curve in the 0.5-2\,keV band for these  observations is
constructed using methods described in Georgakakis et al. (2006).  

\begin{figure}
\plotone{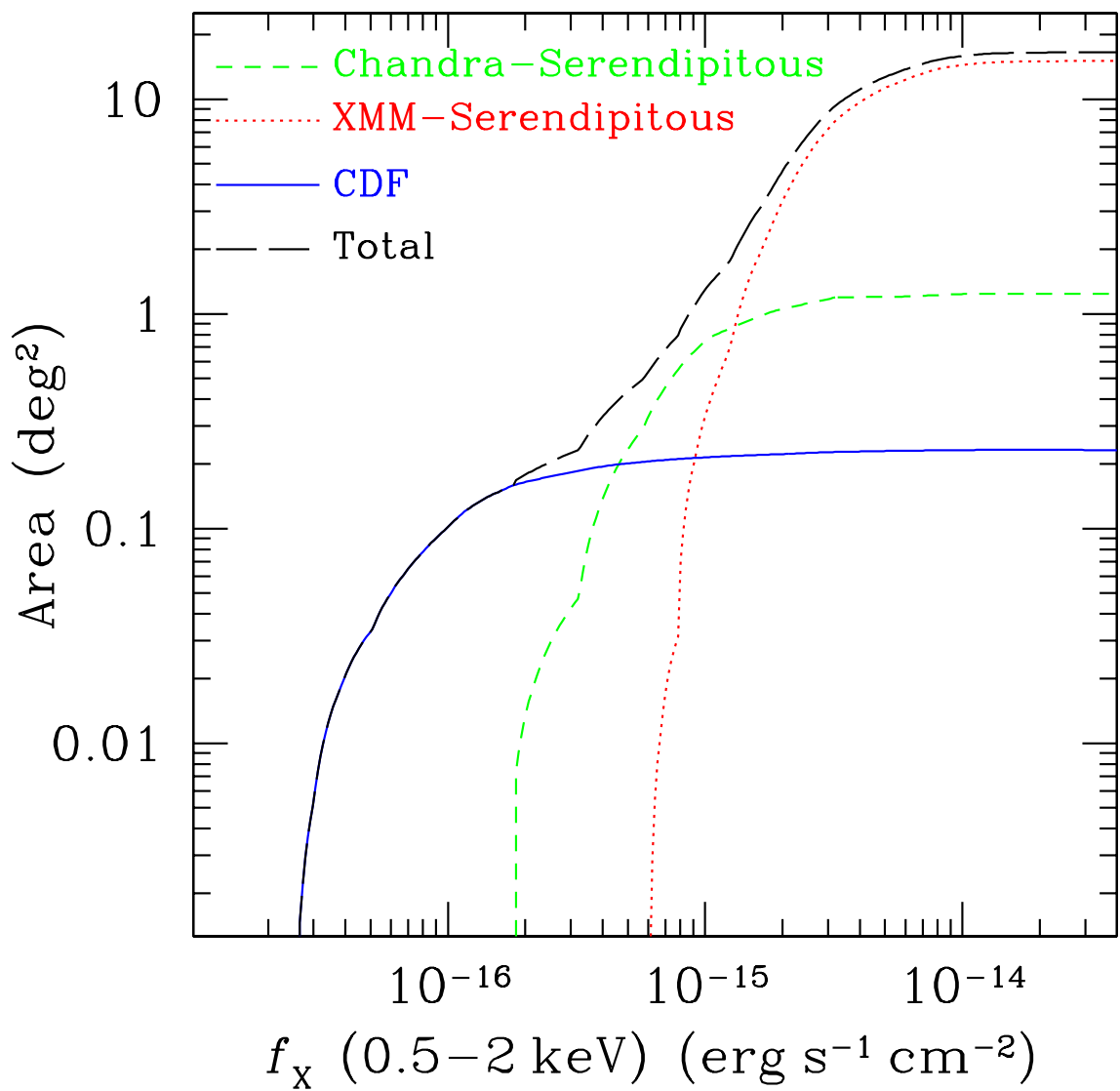}
\caption{
 Solid angle as a function of limiting flux in the 0.5-2\,keV band for
 the X-ray surveys used in this paper to identify normal galaxies. The
 dotted (red) line is for the 1XMM+NHS sample described by Georgakakis
 et al. (2006), the dashed (green) line corresponds to the {\it
 Chandra} serendipitous survey presented in this paper while, the
 continuous (blue) curve is for the combined CDF-North and South. The 
 total sensitivity curve of these independent surveys is plotted with
 the (black) long-dashed line.\label{area} 
}
\end{figure}

The total area of the combined bright sample (1XMM+NHS+XAssist) is
about 15\,deg$^2$ with a sensitivity curve shown in Figure \ref{area}. A
total of 68 normal   galaxy candidates are selected by  searching for
sources with $\log f_X / f_{opt} < -2$, soft X-ray spectral properties
to avoid type-II Seyferts and optical spectra (available for 55 of the
68 sources) consistent with stellar emission rather than AGN activity.
For the $\log f_X / f_{opt}$ calculation we adopt the definition of
Stocke et al. (1991) assuming $\Gamma=1.8$ and 
$B-V=0.8$.

At faint fluxes,  $f_X ( 0.5 - 2.0 \rm \,keV ) \approx 10^{-17} -
10^{-15} \, erg \, s^{-1} \, cm^{-2}$, we use the CDF-North 
and South (Alexander et al. 2003; Giacconi et al. 2002) to identify
normal galaxy  candidates. We focus in particular, on a total of 104
sources (28 in CDF-S and 76 in CDF-N) classified normal galaxies by
Bauer et al. (2004) on the basis of their X-ray and  optical
properties (e.g. hardness ratios, X-ray luminosity, optical emission
lines), which appear consistent with stellar processes. This 
sample  has a less stringent X-ray--to--optical flux ratio cut,  $\log
f_X /  f_{opt}<-1$, compared to the galaxies identified in the shallow
surveys above. Because of evolution and/or differential X-ray/optical
k-corrections, a more stringent X-ray--to--optical flux ratio cut
(e.g. $\log f_X /  f_{opt}= -2$) is likely to affect more seriously
these deeper samples which probe on average higher redshifts,
resulting in incompleteness. On the contrary, the $\log f_X /
f_{opt}$ cut is likely to has little impact in the brighter lower-$z$
samples described  above  (e.g. Georgantopoulos et al. 2005;
Tzanavaris et al. 2006; see section \ref{results}). 

\section{Galaxy classification}\label{classification}
The normal galaxies identified in the surveys above are classified
into early and late types by fitting galaxy templates to their broad
band colours. For the X-ray bright samples presented here we use
primarily the SDSS-DR4 $ugriz$ photometry for spectral fitting. A
total of 12 systems from the 1XMM catalogue do not have SDSS
observations available, in which case we rely on visual inspection of
their optical spectra for classification. For the CDF-South we use the
COMBO-17 photometry (e.g. Wolf et al. 2003) while  for the CDF-North
we also employ the  SDSS-DR4 $ugriz$ photometry. The  exception are 7 
CDF-North systems fainter than the SDSS limit, for which we use the
multiwaveband   observations of Barger et al. (2003) $UBVRIzHK$. 

The template SEDs are constructed by smoothly interpolating between
the 4 galaxy types of Coleman, Wu \& Widman (1980; E/S0, Sbc,  Scd,
Im) extended in the UV and IR wavelengths using the GISSEL98 code 
(e.g. Bruzual \& Charlot 2003). This method provides a total of 60
SEDs from ellipticals (spectral classification 0) to extreme
starbursts (spectral classification 60). Sources with best-fit SEDs in
the range 0--25 are classified early type. This spectral range is in
good agreement with the visual classification of the optical spectra
(e.g. absorption lines), when available. 

Table \ref{tab1} summarizes
the galaxy types found in the four different surveys. It is evident
that the bright 1XMM, NHS and XAssist samples which include only low
redshift galaxies ($z \la 0.2$), consist of roughly equal numbers of early
and late type galaxies. In contrast, the fainter  CDF surveys consist
primarily of late-type galaxies. This suggests that late
and early type galaxies have different evolution with cosmic
time. A quantitative analysis  and a detailed discussion of this point
are presented in the next section.

\begin{table} 
\footnotesize 
\caption{
\rm Number of normal galaxies in the samples used in this paper.
}\label{tab1}  
\begin{center} 
\begin{tabular}{lccc} 
\hline
  sample         & early & late  & all   \\
\hline
 CDF (full)      &  35   & 69    & 104 \\  
 CDF (restricted)&  12   & 22    & 34  \\
 XAssist         &  13   &  9    & 22  \\
 1XMM+NHS$^{\star}$  &  23   & 21   & 46  \\ 
                 &       &      &     \\
 total (full)    &  71   & 99   & 172 \\      
 total (restricted) &  48   & 52   & 102 \\      
\hline

\end{tabular} 
\begin{list}{}{}
\item $^{\star}$2 systems in the 1XMM sample remain unclassified. See
Georgakakis et al. (2006) for details.
\end{list}
\end{center} 
\normalsize  
\end{table}

\begin{figure}
\plotone{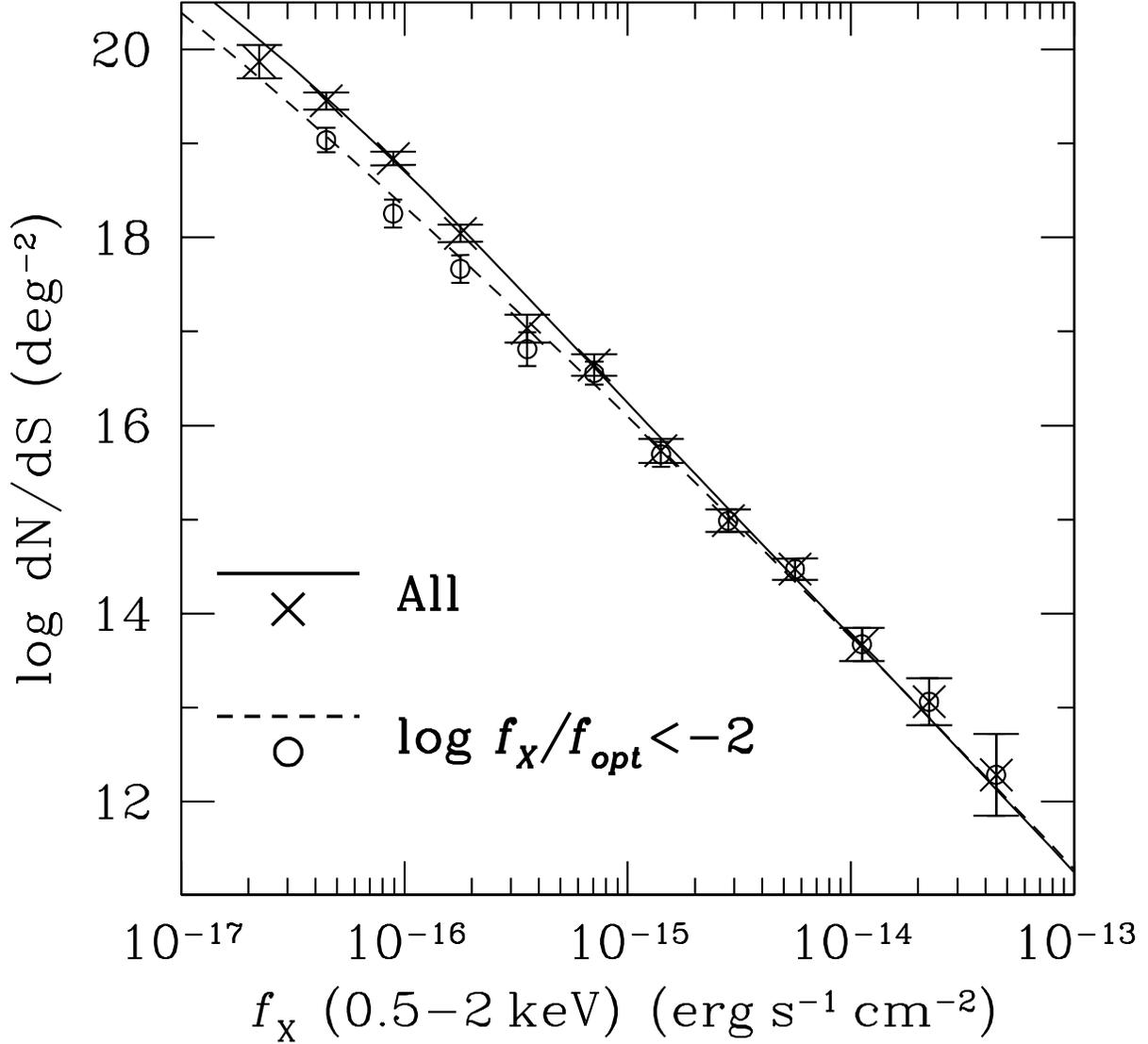}
\caption{
 Differential normal galaxy counts in the 0.5-2\,keV spectral 
 band. The crosses are the combined sample of normal galaxy
 candidates including all systems from the CDF-North and South without
 applying any cutoff in $\log f_X / f_{opt}$. The  continuous line is
 the prediction of the Georgakakis et al. (2006) XLF for
 both early and late type galaxies assuming the maximum likelihood
 evolution of the form $\propto (1+z)^{1.92}$. The open circles
 represent the subsample that only includes CDF galaxies with
 $\log f_X / f_{opt} < -2$. The corresponding dashed line is the no
 evolution prediction ($p=0$) of the Georgakakis et al. (2006)
 XLF.\label{fig_dnds_all}      
 }
\end{figure}

\section{The galaxy X-ray number counts}\label{results} 
In this section we discuss the number density of X-ray selected normal
galaxies over 4 decades of flux in the context of evolution models. We
choose to use the differential galaxy counts, $\rm dN / dS$, instead
of the cumulative $\log N - \log S$, frequently  employed in X-ray
astronomy, because of the independence of individual flux bins
simplifying the estimation of errors and the interpretation of the
results. 

Some contamination from low-luminosity AGN is expected in the
ultra-deep {\it Chandra} fields used here, which comprise, on
average, higher-$z$ systems that are harder to study in
detail. Therefore, the full Bauer et al. (2004) sample used here
(total of 104 galaxies) likely represents close to the upper 
bound in the number density of X-ray selected normal galaxies. We also  
independently  consider galaxies in the Bauer et al. (2004) sample
with low X-ray--to--optical flux ratio, $\log f_X / f_{opt} < -2$. The
AGN contamination in this restricted  subsample is expected to be
minimal, in the expense of possibly missing powerful starbursts or
massive ellipticals, suggested to occupy the  $\log f_X / f_{opt} >
-2$ region of parameter space (e.g. Alexander et al. 2002; Bauer et
al. 2002). 
More recently, Tzanavaris et al. (2006) cross-correlated  the 2dF
Galaxy Redshift Survey with public {\it XMM-Newton} and {\it Chandra} 
observations and found a non-negligible fraction normal galaxies at
$z\approx0.1$, mostly early types however, with  $\log f_X / f_{opt} >
-2$. The evidence above suggests that the Bauer et al. (2004)
CDF-North and South restricted subsample with $\log f_X / f_{opt} <
-2$ is likely to represent  the lower bound in the number density of
normal galaxies.     

Figure \ref{fig_dnds_all} presents the 0.5-2\,keV differential counts
for all normal galaxy candidates in the flux range $f_X ( 0.5 - 2.0
\rm \,keV ) \approx  10^{-17} - 10^{-13} \, erg \, s^{-1} \, cm^{-2}$
identified in the surveys described in  section \ref{sample}. The $\rm
dN/dS$ in this figure is constructed using the sensitivity curve shown
in Figure \ref{area}. We consider separately the two samples that
include (i) all the Bauer et al. (2004) galaxies and (ii) only those
systems in the CDF-North and South with   $\log f_X / f_{opt} <
-2$. The former is hereafter referred to as the full sample while, the
latter is dubbed the restricted sample. We fit the unbinned data for
these two galaxy samples with a power-law using the parametric Maximum
Likelihood method.  This yields slopes of $-2.47\pm0.04$ and
$-2.22\pm0.05$  for the full and restricted samples respectively.
Using the full sample and adopting the combined early and late-type
X-ray luminosity function (XLF) of Georgakakis et al. (2006) assuming
luminosity evolution of the form $(1+z)^p$ we find
$p=1.92\pm0.35$. This is consistent with the evolution derived by 
Norman et al. (2004) and Ranalli et al. (2005). The $\rm dN/dS$ of the
restricted sample with $\log f_X / f_{opt} < -2$ is best fit with
negative values of $p$, implying no-evolution. The evidence above
suggests that in the case of the total sample (both early and late
types) the  evolution is driven by the $\log f_X / f_{opt} > -2 $
systems.   
 
\begin{figure}
\plotone{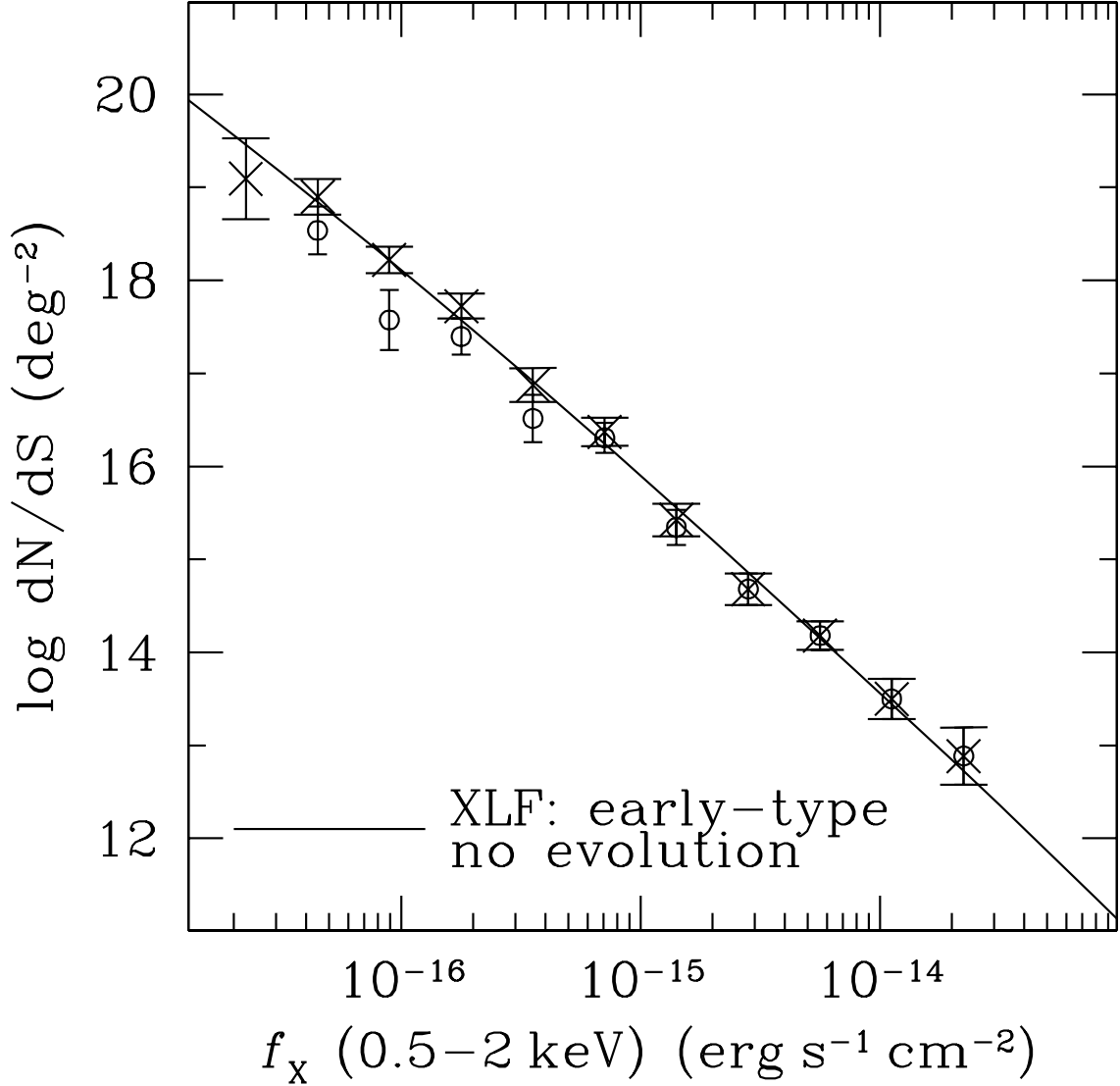}
\caption{
 Differential counts for early-type normal galaxies in the
 0.5-2\,keV spectral band. The crosses are evolved galaxy candidates,
 including all systems from the CDF-North and South without applying
 any cutoff in $\log f_X /  f_{opt}$. The open circles represent the
 subsample that only includes CDF early-type galaxies with $\log f_X /
 f_{opt} < -2$. The  continuous line is the prediction of the Georgakakis
 et al. (2006) XLF for early-type galaxies assuming no evolution. 
}\label{fig_dnds_abs}    
\end{figure}

\begin{figure}
\plotone{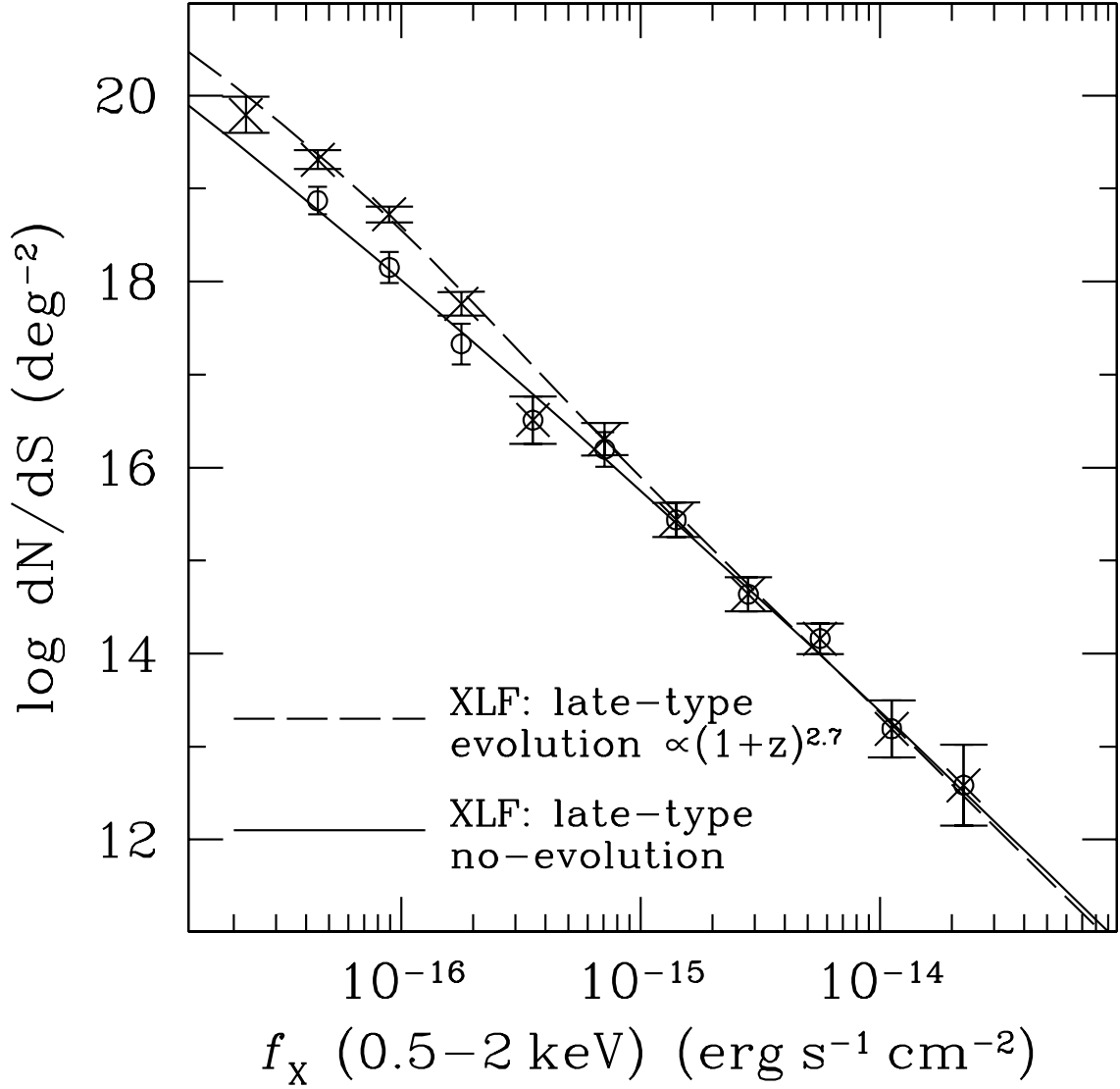}
\caption{
 Differential counts for late-type normal galaxies in the
 0.5-2\,keV spectral band. The symbols are as in Figure
 \ref{fig_dnds_abs}. The  continuous and dashed lines are the
 predictions of the Georgakakis et al. (2006) late-type
 galaxy XLF for luminosity evolution $\propto (1+z)^{2.7}$ (i.e. the
 maximum likelihood fit to the full subsample) and no-evolution
 respectively.    
}\label{fig_dnds_em}    
\end{figure}

In addition to the total galaxy counts we also consider early and
late-type galaxy subsamples independently. Figure \ref{fig_dnds_abs} 
plots the $\rm dN/dS$ for evolved galaxies according to the
classification presented in section \ref{classification}. The full and
restricted subsamples for early-type galaxies are plotted separately,
delineating the upper and lower X-ray number density limit of evolved
galaxies. In the case of the full subsample, using the parametric
maximum likelihood method and assuming luminosity evolution of the
form $\propto (1+z)^{p}$ for the Georgakakis et   al. (2006)
early-type XLF we constrain $p = 0.59^{+0.78}_{-0.59}$ consistent
with no evolution. The restricted subsample requires negative values
of $p$ implying either incompleteness of the CDFs  due to the $\log f_X /
f_{opt} < -2$ cutoff or uncertainties in the early/late type
classification. Nevertheless the evidence above suggests that optically
classified early-type galaxies show little (if any) evolution at X-ray
wavelengths, at least out to  $z \approx 0.5$, the median redshift of
evolved galaxies in the CDFs.    

In Figure \ref{fig_dnds_em} we plot the differential counts for
late-type galaxies. These are compared with the $\rm dN/dS$ estimated
by integrating the late-type galaxy XLF of Georgakakis et al. (2006)
assuming different scenarios for the evolution of this
population. The restricted subsample, comprising systems with $\log
f_X / f_{opt} < -2$, appears in agreement with the no evolution 
model. Indeed, a maximum likelihood fit to the data, parameterising
the evolution as $\propto (1+z)^{p}$ for the Georgakakis et al. (2006)
late-type XLF, yields $p = 0.41^{+0.85}_{-0.41}$ consistent with no
evolution. On the contrary, a similar analysis for the full subsample
gives $p = 2.70 \pm 0.35$, suggesting evolution out to $z\approx0.4$,
the median redshift of the  late-type subsample. This suggests that it
is the $\log f_X / f_{opt} > -2$ galaxy population that evolves the
faster. Indeed, these sources  are suggested to comprise a large
fraction of starburst galaxies expected to show most dramatic
evolution with redshift (e.g. Alexander et al. 2002).  The evolution
derived here for the late-type sample is comparable within the
uncertainties to this of the full sample. This is not surprising as
most of the galaxies at high redshift, which drive the evolution, are
late-types. The derived evolution at X-ray wavelengths is consistent
with that estimated for star-forming galaxy samples selected at other
wavelengths (e.g. Hopkins 2004). Therefore, the late-type X-ray
population studied here closely maps the star-formation history of the
Universe, at least out to the median redshift of the sample, $z \approx
0.4$. The evolution rate derived here, $p = 2.70 \pm 0.35$, is
consistent with those models of Ghosh \& White (2001) that adopt a
slow evolution timescale for the low-mass X-ray binary population,
$\ga 1$\,Gyr. Shorter timescales predict exponents much higher than
that derived here, almost independent of the star-formation history
of the Universe at higher redshift.

Finally, we caution that low-luminosity AGN contamination may be
present in our sample. This problem is likely to be more severe
for higher-$z$ systems, where weak AGN optical signatures are harder
to identify against the stellar continuum (e.g. Severgnini et
al. 2003). Although, the presence of low-luminosity AGN may bias our
results to higher values for the evolution parameter, $p$, we argue
that this is unlikely to modify the main conclusions presented here. 
The X-ray spectra of many low luminosity AGN are indeed, dominated by
thermal emission below about 2\,keV (Levenson, Weaver \& Heckman 2001;
Terashima et al. 2002), suggesting stellar  processes. Even if the
late-type galaxy sample used in this paper has some level of residual
AGN contamination, this is likely to have only a minor contribution to
the 0.5-2\,keV, employed here.  


\section{Acknowledgments}
This work has been supported by the program `Studying Galaxies with
NASA's mission {\it Chandra}' of the Hellenic General Secretariat for
Research and Technology. We acknowledge use of data from the Sloan
Digital Sky Survey and the XAssist {\it Chandra} source
catalogue. This work has used data from the CXC data archive.


\begin{thebibliography}{} 
\bibitem{} Alexander D., et al., 2003, AJ, 126, 539

\bibitem{} Alexander D. M., Aussel H., Bauer F. E., Brandt W. N.,
Hornschemeier A. E., Vignali C., Garmire G. P., Schneider D. P., 2002,
ApJ, 568L, 85 





\bibitem{} Bauer, F.E., Alexander, D.M., Brandt, W.N., Schneider,
D.P., Treister, E.,  Hornschemeier, A.E., Garmire, G.P., 2004, AJ,
128, 2048 


\bibitem{} Brandt W. N., et al., 2001, AJ, 122, 1

\bibitem{} Bruzual G. \& Charlot S.,  2003, MNRAS, 344, 1000











\bibitem{} Georgakakis, A., Chavushyan V., Plionis M.,
Georgantopoulos, I., Koulouridis E., Leonidaki I., Mercado A., 2006,
MNRAS, in press (astro-ph://0601212)

\bibitem{} Georgakakis, A.E., Georgantopoulos, I., Basilakos, S., 
Plionis, M., Kolokotronis, V., 2004, MNRAS, 354, 123 




\bibitem{} Georgantopoulos I., Georgakakis A., Koulouridis E., 2005,
MNRAS, 360, 782

\bibitem{} Ghosh P. \& White N. E., 2001, ApJ, 559, 97L

\bibitem{} Giacconi R., et al.,  2002, ApJS, 139, 369



\bibitem{} Hopkins, A., 2004, ApJ, 615, 209


\bibitem{} Hornschemeier A. E. et al., 2003, AJ, 126, 575

\bibitem{} Hornschemeier A. E., Brandt W. N., Alexander D. M., Bauer
F. E., Garmire G. P., Schneider D. P., Bautz M. W., Chartas G., 2002,
ApJ, 568, 82.










\bibitem{} Levenson N. A., Weaver K. A., Heckman T. M., 2001, ApJ, 550, 230








\bibitem{} Norman C., et al., 2004, ApJ, 607, 721 

\bibitem{} Ptak A., \& Griffiths R., 2003, ASPC, 295, 465




\bibitem{} Ranalli, P., Comastri, A. Setti, G., 2005, A\&A, 440, 23



\bibitem{}Severgnini P. et al., 2003, A\&A, 406, 483



\bibitem{} Schneider D., et al., 2005, AJ, 130, 367





\bibitem{} Stocke, J.T. et al., 1991, 76, 813  






\bibitem{} Terashima Y., Iyomoto N., Ho L. C., Ptak A. F., 2002, ApJS,
139, 1





\bibitem{} Tzanavaris P., Georgantopoulos I., Georgakakis A., 2006, in preparation.




\bibitem{} Zezas A., 2000, PhD Thesis, University of Leceister




\end{thebibliography}
\end{document}